\documentclass{elsart}
\usepackage{latexsym}
\usepackage{epsfig}
\usepackage{natbib}

\def\url#1{#1} 

\begin{document}

\begin{frontmatter}

\title{Spectral analysis of extinguished sunlight}

\author{Fr\'ed\'eric \snm Zagury\thanksref{email1}}
\address{02210 Saint R\'emy Blanzy, France}
\thanks[email1]{E-mail: fzagury@wanadoo.fr}
\and
\author{Florence \snm Goutail\thanksref{email2}}
\address{Service d'A\'eronomie, CNRS,             
91371 Verri\`eres-le-Buisson, France }
\thanks[email2]{E-mail: Florence.Goutail@aerov.jussieu.fr}
\received{December 2002}

 \begin{abstract}
SAOZ (Syst\`eme d'Analyse par Observation Z\'enitale)  is a balloon born 
experiment which determines the column 
density of several molecular species from the 
visible spectrum of sunlight.
We will use sequence of spectra collected during a sunset 
to discuss atmospheric extinction,
and the nature of the radiation field in the atmosphere.
    
The radiation field in the atmosphere is, from daylight to sunset, 
and with a clear sky, dominated by light coming from the direction of the sun. 
This light is composed of direct sunlight (extinguished by the gas),
and of sunlight forward-scattered by aerosols.
As the sun sets, aerosol scattering is first perceived 
towards  the UV.
It progressively replaces direct sunlight over all of the spectrum.
Our analysis permits fixing the main parameters of each 
component of the radiation field at any time.

The fits we find for the extinction of sunlight in the atmosphere 
must also apply to starlight.
Thus, the present work can be used in astronomy to correct ground-based spectral 
observations for extinction in the atmosphere.
   \end{abstract} 
 \begin{keyword}
{atmospheric effects; diffusion; scattering; radiative transfer}
     \PACS
42.68.J  \sep    
42.68.A,  \sep
94.10.G  \sep	  
92.60  \sep	  
03.80  \sep
94.10.L  \sep
92.60.E  \sep
51.20  \sep
95.30.Jx 
  \end{keyword}   
\end{frontmatter}
 \section{Introduction} \label{intro}
SAOZ is a balloon experiment in which the light collected by a wide 
field conical mirror is analyzed in the visible by two large bandwidth spectrometers.
The original aim of SAOZ is to collect informations on the column 
densities of several molecular species (ozone, nitrogen, water vapor) in 
the atmosphere.

The spectra acquired by the spectrometers can 
also be used to analyze the radiation field at the balloon's place, in 
the visible wavelength domain.

In this paper we consider a sequence of spectra covering a 
sunset in May 2002 (section~\ref{data}).
Fits of the spectra will be used to analyze the extinction of sunlight through 
successive, and thicker, layers of the atmosphere (sections~\ref{ana} and 
\ref{dis}), to derive a general analytical expression for the radiation field in 
the atmosphere.  
This will further help in understanding what the radiation field 
transmitted through a clear atmosphere, by a background source of light (as the sun), 
should be.

The pertinence of these conclusions, reached through the sole analysis of 
extinguished sunlight spectra, can be verified 
-because of the proximity of the sun, and of our knowledge of the atmosphere- 
giving weight to the idea that a 
formal, and model-independent study of spectra can lead to
necessary conclusions on the nature of the media crossed 
by light (precisely; on the particles which compose these media).

In section~\ref{astro} we discuss possible applications of this work 
in a 
fundamental problem of ground-based astronomy: the correction for
atmospheric extinction. 
\section{data} \label{data}
\begin{table*}[h]
 \caption[]{Parameters for the spectra used in the article}		
      \[
    \begin{tabular}{|c|ccc|c|c|c|c|ccc|}
\hline
n$^{\circ}$ & \multicolumn{3}{|c|}{U.T.=L.T.-2h}& SZA $\,^{\,(1)}$ & Tg. Alt. $\,^{\,(2)}$ & $N_{O_{3}} 
\,^{\,(3)}$ & $c_1 \,^{\,(4)}$ && Fit &\\ 
& h. & min. & sec. & $^{\circ}$ & m &cm$^{-2}$&& $c_2 \,^{\,(5)}$ & $c_3 \,^{\,(5)}$ & $c_4 
\,^{\,(5)}$ \\   \hline
1 &  19 &  08 &  51 & $89.69$ & $29094.5$ & $ 9.35\,10^{18} \pm  8\,10^{16}$ & $  1.0$ & $0.0005$ & $0.00$ & $0.000$ \\
 2 &  19 &  18 &  6 & $91.20$ & $27553.2$ & $ 8.22\,10^{19} \pm  3\,10^{17}$ & $  1.0$ & $0.0075$ & $0.00$ & $0.000$ \\
 3 &  19 &  27 &  14 & $92.68$ & $22355.4$ & $ 2.23\,10^{20} \pm  6\,10^{17}$ & $  1.0$ & $0.029$ & $0.00$ & $0.000$ \\
 4 &  19 &  34 &  56 & $93.91$ & $15930.8$ & $ 2.62\,10^{20} \pm  7\,10^{17}$ & $  1.5$ & $0.094$ & $0.00$ & $0.004$ \\
 5 &  19 &  38 &  56 & $94.53$ & $12437.2$ & $ 2.16\,10^{20} \pm  3\,10^{18}$ & $  3.5$ & $0.16$ & $0.00$ & $0.003$ \\
 6 &  19 &  39 &  43 & $94.66$ & $11654.8$ & $ 2.23\,10^{20} \pm  4\,10^{18}$ & $  5.3$ & $0.165$ & $0.00$ & $0.005$ \\
 7 &  19 &  40 &  44 & $94.82$ & $10696.8$ & $ 3.12\,10^{20} \pm  1\,10^{19}$ & $ 13$ & $0.17$ & $0.60$ & $0.068$ \\
 8 &  19 &  47 &  31 & $95.86$ & $ 4776.1$ & $ 3.72\,10^{20} \pm  2\,10^{20}$ & $  5$ & $0.5$ & $1.04$ & $0.021$ \\
 9 &  19 &  51 &  46 & $96.51$ & - & $ \sim 3.35\,10^{20}$ & $100$ & $0.4$ & $1.40$ & $1.933$ \\
 10 &  19 &  58 &  46 & $97.56$ & - & $ \sim 3.00\,10^{20}$ & $250$ & $0.5$ & $1.03$ & $0.829$ \\
 11 &  20 &  02 &  27 & $98.10$ & - & $ \sim 4.00\,10^{20}$ & $800$ & Inf & $1.35$ & - 
\\
\hline
\end{tabular}    
    \]
\begin{list}{}{}
\item[$1$] Zenital angle of the sun. 
\item[$2$] Tangent altitude of the trajectory of sunrays, calculated 
from the respective positions of the balloon and of the sun.
The calculation takes into account the bending of sunrays in the 
atmosphere. 
\item[$3$] Ozone column density with observational error margin. For 
the last three spectra we have no 
reliable observational measure of the column density (see text): 
$N_{O_{3}}$ was estimated by hand from the spectra themselves. 
No error estimate on $N_{O_{3}}$ is given for these spectra.
\item[$4$] Scaling factor used in the figures, for each spectrum. 
\item[$5$] The fit of the spectra are proportional to: 
$(e^{-c_2/\lambda^{4}}+c_4\lambda^{-1})e^{-c_3/\lambda}$
\end{list}
\label{tbl:spec}
\end{table*}
\subsection{Instrument} \label{saoz}
Up until recently, the balloon-borne SAOZ instrument consisted of a 
lightweight solar occultation spectrometer, 
in which, the solar tracker is a conical mirror defining a field of 
view of $[-5^{\circ}¥,\, +10^{\circ}]$ elevation and $360^{\circ}$ 
azimuth (see \citet{saoz1,saoz2,saoz3} for a complete description; 
some 
informations on SAOZ are also available at:
http://www.aero.jussieu.fr/themes/CA/).
This instrument performs measurements from $3000\,\rm\AA$ to $6000\,\rm\AA$.

In this standard configuration, SAOZ is used to measure the 
stratospheric column densities -between the sun and the balloon- 
of ozone and nitrogen.
The ozone measure is precise to more than 
$5\,\rm\%$ if the light received by the 
receptor in the ozone wavelength range is direct sunlight alone.
When the sun is low in the horizon direct sunlight diminishes and the receptor, because of its' 
large azimutal aperture, receives a large amount of scattered 
sunlight from different directions; in this case the ozone measure becomes meaningless.
The relative contribution of scattered light to the radiation field 
at the balloon location is monitored by means of the color index. 

Pressure, temperature and GPS (Global Positioning System) altitude 
and location are also measured on-board 
with an accuracy of respectively $1\,$hPa, $0.5\,^{\circ}$K and 
$100\,$m. 

A new spectrometer is now added to the experiment to measure 
water vapor in the stratosphere.
This spectrometer covers the $[0.4\,\rm\mu m, \, 1\,\rm\mu m]$ wavelength domain with a resolution 
of $6.33\,\rm\AA$.
\subsection{Data} \label{sp}
The spectra presented in the paper come from the first 
test-flight with the new spectrometer.
SAOZ was launched from Aire sur Adour (Landes, France) on May 14, 2002, 
and stabilized (to within $50\,$m during the flight) at an altitude of $\sim 28900\,\rm m$.

Spectra were alternatively taken by each spectrometer, at an average 
rate of one spectrum per $50$ seconds.
The old spectrometer is used to determine the column 
density of ozone.
For studying the radiation field at the balloon's location, we
used data from the new spectrometer, because of its larger wavelength coverage.

We used the observations from a sun at zenital angle 
$SZA=89.7$ (the sun 
is observed through a thin stratospheric layer), to  
$SZA=98.1$ (i.e. the sun is below the horizon).
\subsection{Ozone correction} \label{ozone}
\begin{figure}[h]
\resizebox{!}{\columnwidth}{\includegraphics{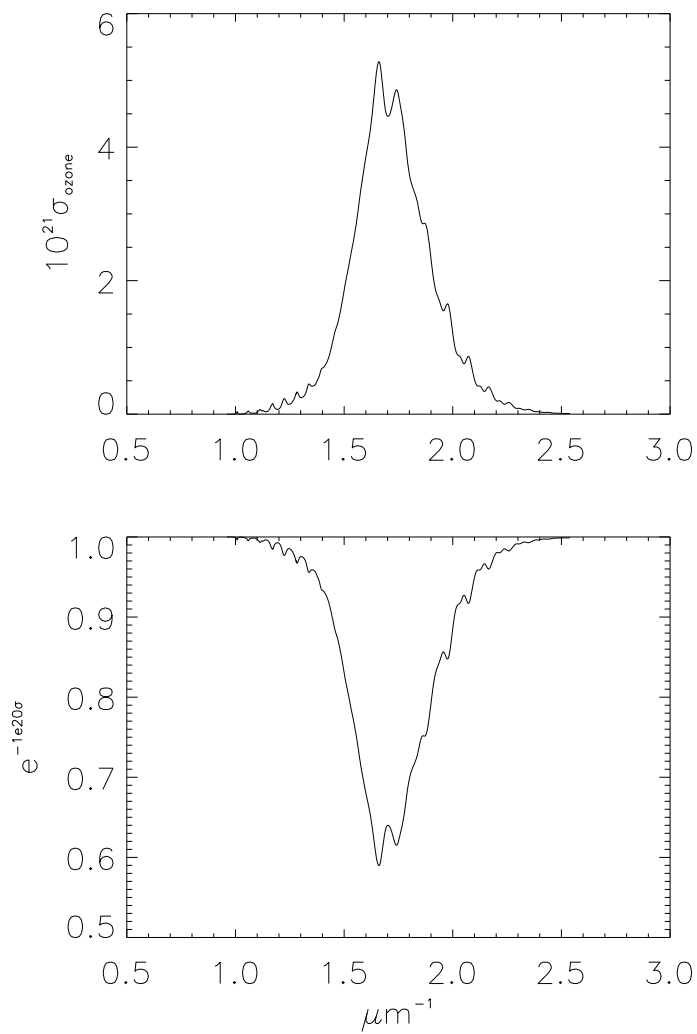}} 
\caption{
\emph{Top:} profile of the ozone absorption cross-section.
\emph{Bottom:} ozone absorption for an ozone column density of 
$N_{O_{3}¥}=10^{20}\,\rm cm^{-2}$.
} 
\label{fig:ozone}
\end{figure}

The broad-band ozone absorption occupies the central part of 
the spectra, between $\sim 5000\,\rm\AA$ and $\sim 7000\,\rm\AA$.
The wavelength dependent absorption cross section of ozone, 
$\sigma_{\lambda}$, is plotted in Figure~\ref{fig:ozone}.
To correct a spectrum for the absorption by ozone, the spectrum needs to be 
multiplied by $e^{\sigma_{\lambda}¥ N_{O_{3}¥}¥}$, with $N_{O_{3}}$ the column density of ozone.
\subsection{Table~\ref{tbl:spec}} \label{tbl}
Table~\ref{tbl:spec} summarises the informations on the spectra.

The first and second columns are the number given to each spectrum 
and the universal time of the observation (local time=U.T.+02h).

Columns~3 and 4 are the angular distance of the 
sun to the zenith (SZA), and the tangent altitude of the 
average trajectories of sunrays.
The latter takes into account the bending of sunrays due to 
the  variations of the refractive index of the atmosphere.
It is used to define the different layers crossed by 
sunlight in the atmosphere.

The $N_{O_{3}}$ column is the 
ozone column density for spectrum~(1), to spectrum~(7),  measured by 
the old spectrometer (section~\ref{saoz}).
The error on  $N_{O_{3}}$ increases with time due to the 
setting of the sun and the growing importance of scattered sunlight.
For the four last spectra  $N_{O_{3}}$ was estimated 
from the best correction applied to the spectra to reduce, or 
suppress, the ozone depression (section~\ref{ana}).

The last four columns are the parameters we found for the fits of the spectra.
\section{Analysis} \label{ana}
\begin{figure*}[p]
\resizebox{!}{1.5\columnwidth}{\includegraphics{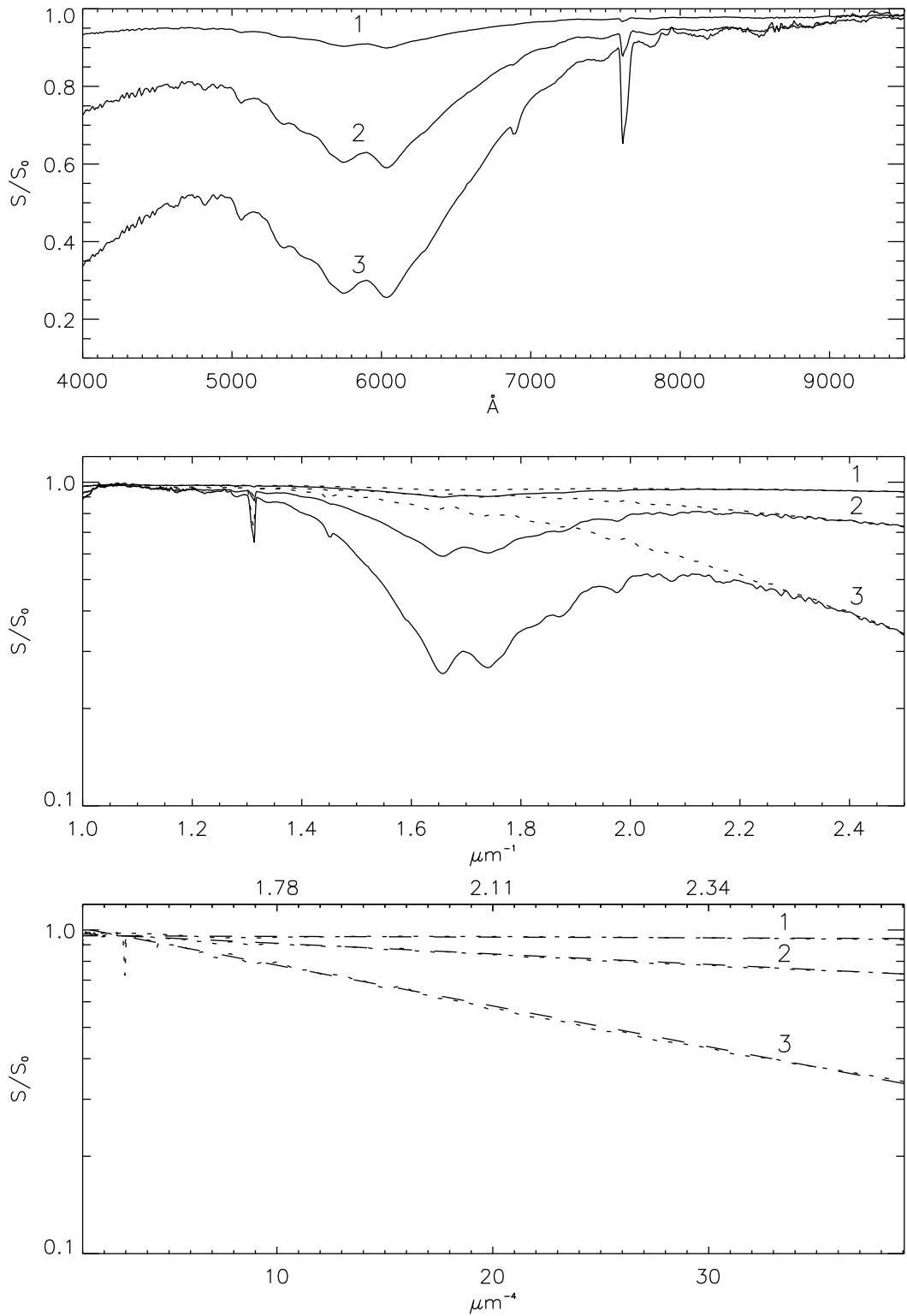}} 
\caption{First spectra of the sun observed through the stratosphere.
Spectra are labeled by their number in Table~\ref{tbl:spec}.
\emph{Top:} the spectra are divided by the spectrum of the sun and 
are scaled to 1 in the near-infrared, x-axis 
in $\rm\AA$.
\emph{Middle:} same spectra but presented against $1/\lambda$, and 
with 
the correction for absorption by ozone (in dots).
\emph{Bottom:} the spectra (in dots) corrected for ozone extinction 
decrease 
exponentially as $1/\lambda^{4}$ (exponential fit in dashes). 
The bottom x-axis is $1/\lambda^{4}$.
$1/\lambda$ is represented on the top x-axis of the plot.
} 
\label{fig:ext1}
\end{figure*}

The following sections analyze the sequence of decreasing 
sunlight spectra from a sun above the atmosphere, to the complete extinction 
of sunlight below the horizon.
We distinguish four periods.
For each of these periods we have extracted a few representative 
spectra.
All the spectra are normalized by a spectrum of the sun at $19^{h}03^{m}43^{s}$, 
when the sun is still observed above the atmosphere.
The spectra are scaled to 1 in the near-infrared ($\lambda\sim 
9500\,\rm\AA$).
The scaling factor is $c_{1}$, sixth column of Table~\ref{tbl:spec}. 
\subsection{Direct sunlight alone (figure~\ref{fig:ext1})} 
\label{direct}
Top of Figure~\ref{fig:ext1} is a plot of the 
first normalised spectra of the sunset against wavelength.
While the balloon had, a few minutes ago, been observing the sun above the 
atmosphere, sunrays now cross the upper atmospheric layer, the 
stratosphere.

The large absorption band between $5000\,\rm\AA$ and $7000\,\rm\AA$ 
is due to ozone (the Chappuis visible bands).
Ozone absorption increases with time, with 
the increase of optical path followed  by sunrays through the stratosphere.

The middle plot of the figure reproduces the spectra of the upper 
plot (plain lines), as a function of wave-number.
The dotted lines are the same spectra after correction for 
$O_{3}$ absorption (see section~\ref{ozone}).

\begin{figure*}[p]
\resizebox{!}{1.5\columnwidth}{\includegraphics{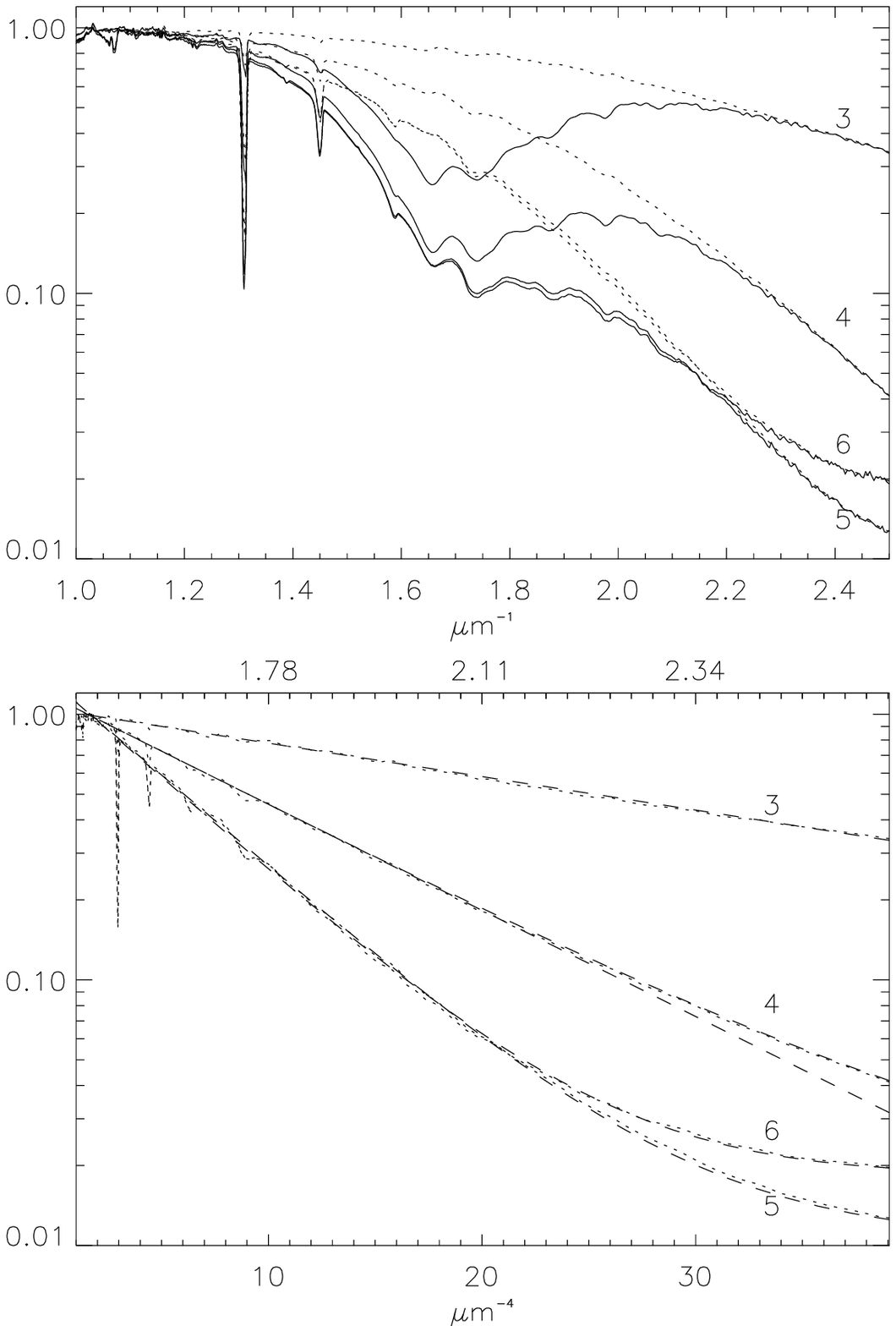}} 
\caption{
Spectrum (3) from figure~\ref{fig:ext1}, and spectra (4) and (5) 
(Table~\ref{tbl:spec}). 
Original spectra are in solid lines, spectra corrected for 
ozone absorption in dots, fits in dashes.
\emph{Bottom:} The $1/\lambda^{4}$ gas 
extinction increases with time. Towards the UV, this extinction is no 
longer enough to 
fit the spectra (see spectrum (4)).
The exact fits comprise an additional term,
$\propto \lambda^{-1}¥$, corresponding to low optical depth 
scattering of sunlight by aerosols 
in the troposphere.
} 
\label{fig:ext2}
\end{figure*}

The spectra corrected for the broad-band ozone absorption 
all decrease as an exponential of $1/\lambda^{4}$ (bottom plot of figure~\ref{fig:ext1}).
We conclude that extinction through the stratosphere is of Rayleigh type, due to  the 
stratospheric gas (nitrogen essentially). 
\subsection{Direct sunlight and low optical 
depth scattered sunlight (figure~\ref{fig:ext2})} \label{lowsca}
From $\sim$19$^{h}$:34$^{m}$ on (between spectra (3) and (4)), 
the spectra can no longer be fitted by the Rayleigh extinction alone.
As illustrated by spectrum (4) (figure~\ref{fig:ext2}, bottom 
plot), the $e^{-c_{2}/\lambda^{4}}$ fit still holds for the shortest 
wave-numbers but, towards the UV, the spectra have 
larger values than expected.
The reason for this excess is the addition of a 
component of scattered 
sunlight which appears with the increase of extinction.
As expected, scattered light is first detected towards the UV, where extinction is 
the highest.

The scattered light behaves as $\lambda^{-1}$ (figure~\ref{fig:ext2}, bottom), 
not as $\lambda^{-4}$ as it would if scattering was due to molecules.
We know from scattering theory that the particles, atmospheric aerosols 
in the present case, must be large compared with the wavelength.

The normalized spectra corrected for ozone absorption are now the sum of a 
direct sunlight component, $\propto e^{-c_{2}/\lambda^{4}}$, and a scattered 
sunlight component, $\propto$~$\lambda^{-1}$.
We observe the short-wavenumber rise of scattering: 
the $\lambda^{-1}$ dependence of the scattered sunlight 
corresponds to the small optical depth approximation.

Spectra (5) and (6) have nearly the same gas extinction while the 
scattered components are clearly separated, meaning that
gas extinction and aerosols scattering are not necessarily dependent.

\begin{figure*}[p]
\resizebox{!}{1.5\columnwidth}{\includegraphics{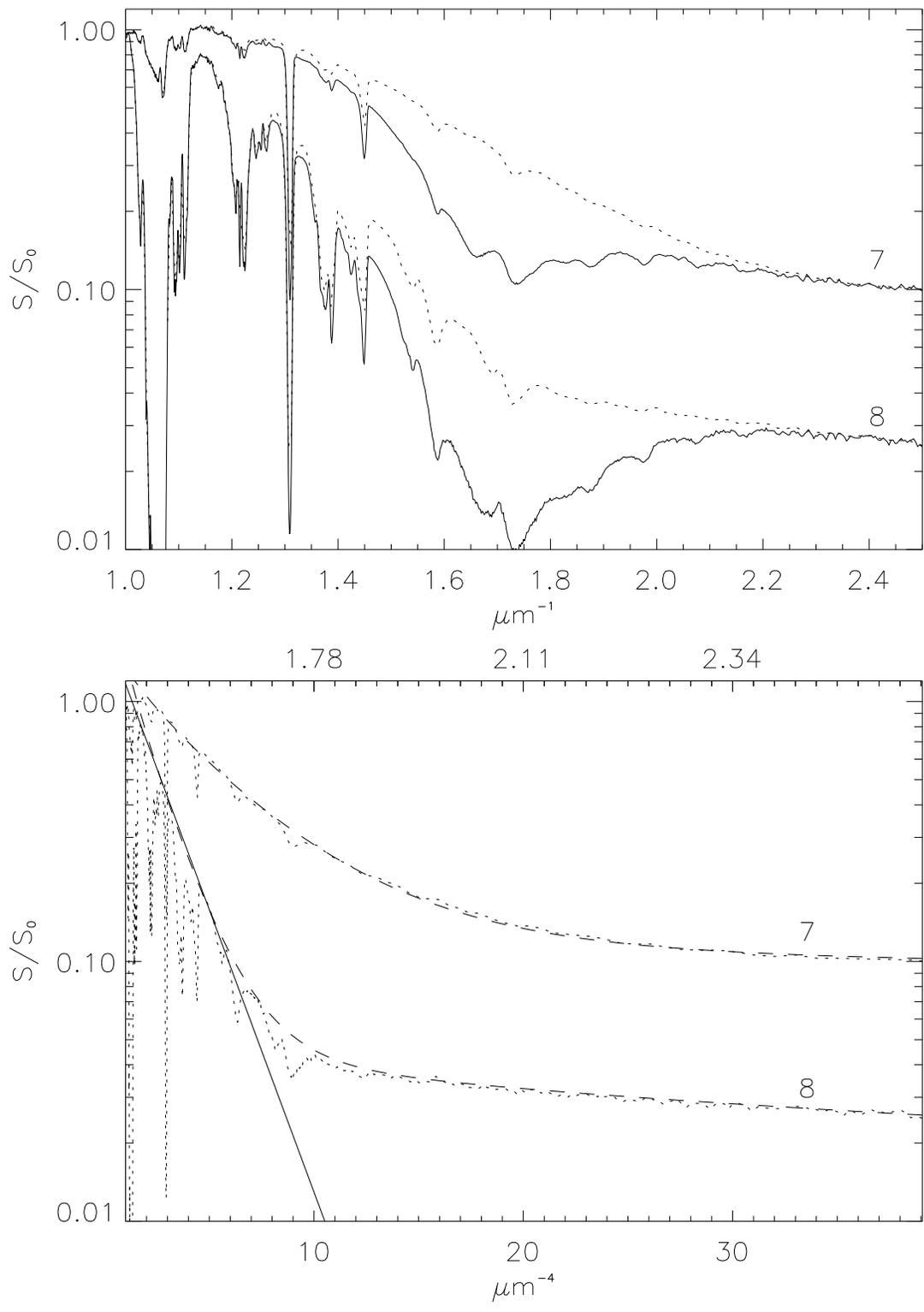}} 
\caption{
Original spectra are in solid line, spectra corrected for 
ozone absorption in dots, fits in dashes.
\emph{Top:} For spectrum (8), the ozone correction is less precise 
than in the preceding spectra, probably due to the increased 
importance of the 
scattered light.
\emph{Bottom:} Fit of the spectra with a direct sunlight component 
and a scattered light component $\propto \lambda^{-1}¥ 
\,e^{-c_{3}¥/\lambda}$.
Direct sunlight extinction is nearly all extinction by the gas,  
$e^{-c_{2}/\lambda^{4}}$.
} 
\label{fig:ext3}
\end{figure*}

The detection of a small column density of aerosols on the path of 
sunrays indicates that 
sunlight now crosses the upper part of the troposphere.
This is confirmed by the independent information on the 
relative position of the balloon and of the sun.
\subsection{Increased importance of the scattered sunlight
(figure~\ref{fig:ext3})} \label{highsca}
After 19$^{h}$:40$^{m}$ the small optical depth approximation no longer 
holds in the observational wavelength range. 
Scattered sunlight is now fitted  by a more general function $\propto 
\lambda^{-1}¥ \,e^{-c_{3}¥/\lambda}$, expected for forward scattering 
\citep{o1spec}.

Direct sunlight rapidly decreases. 
The main cause of direct sunlight extinction is still the 
$e^{-c_{2}¥/\lambda^{4}}$ Rayleigh extinction.
The extinction of direct sunlight by aerosols is comparatively 
negligible ($c_{3}/\lambda \ll c_{2}/\lambda^{4}$).

Observational determination of ozone column density 
becomes meaningless with the increase of scattered sunlight 
(section~\ref{saoz}).
For spectrum (8), scattered light occupies the whole spectrum for 
$1/\lambda>1.7\,\rm\mu m^{-1}$, and the ozone correction had to be adjusted by hand.

\begin{figure*}[p]
\resizebox{!}{1.5\columnwidth}{\includegraphics{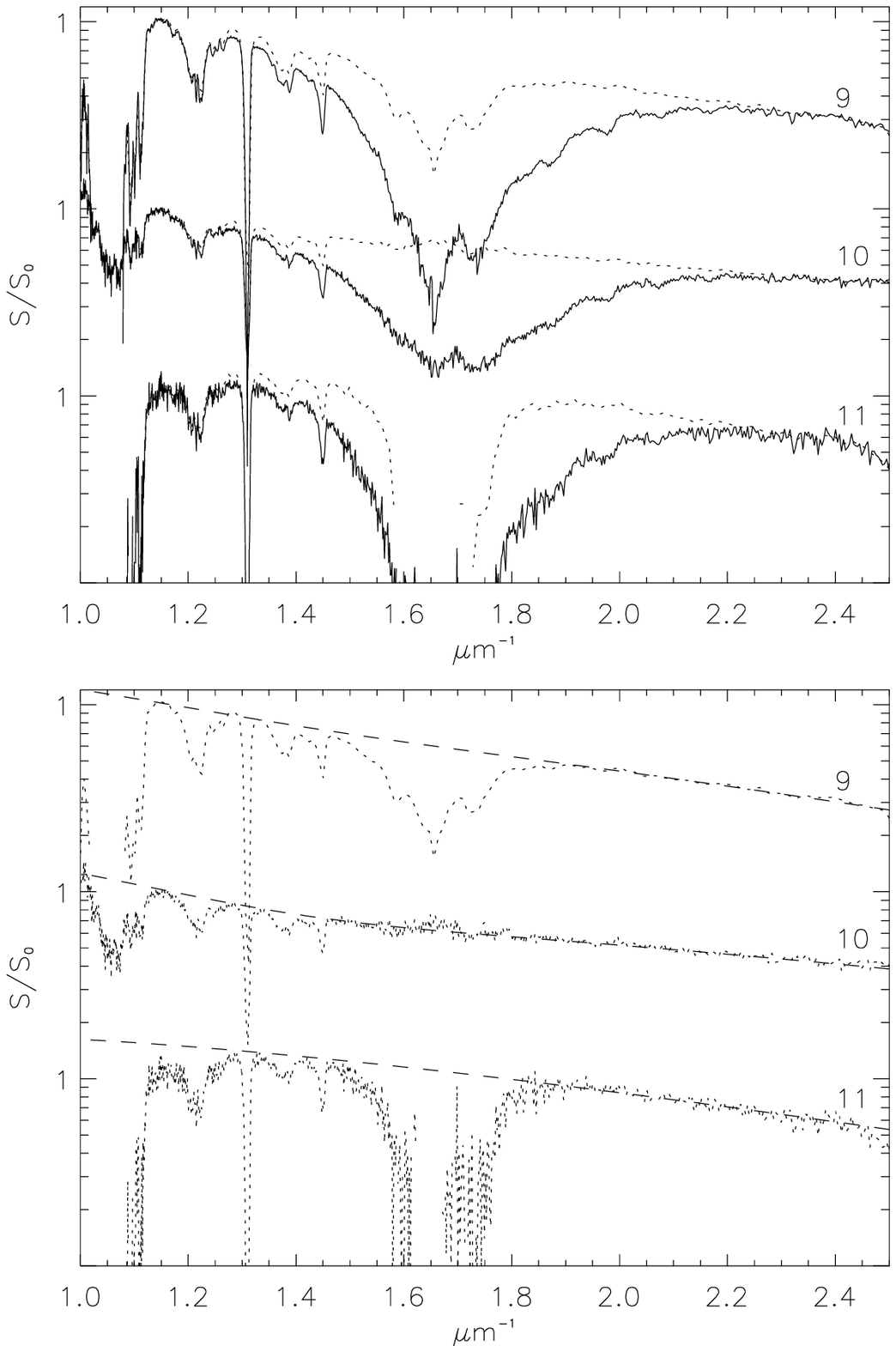}} 
\caption{
Original spectra are in solid line, spectra corrected for 
ozone absorption in dots, fits in dashes.
Direct sunlight is rejected to the near-infrared.
Compared to daylight the radiation field diminishes by 2-3 orders of 
magnitude (column $c_{1}$ in 
Table~\ref{tbl:spec}).
Correction for ozone absorption was adjusted by hand and could not be 
achieved for spectra (9) and (11).
} 
\label{fig:ext4}
\end{figure*}

Scattered light progressively replaces direct sunlight as the sun sets.
\subsection{Scattered sunlight (figure~\ref{fig:ext4})} \label{sca}
When the sun disappears below the horizon scattered light 
occupies all the spectrum except for its nearest infrared part.
Replacement of direct sunlight by scattered 
light is accompanied by an important diminution of the overall 
intensity of the radiation field (2-3 orders of magnitude, column 
$c_{1}$ in Table~\ref{tbl:spec}).

We tried to determine an average $N_{O_{3}¥}$ by correcting the ozone deficit feature
of each spectrum.
The correction found for spectrum (10) leads to 
$N_{O_{3}¥}=3\,10^{20}\,\rm cm^{-2}$.
The ozone absorption feature in spectra (9) and (11) is much deeper than in any 
of the previous spectrum.
The correction was limited to the wings of the absorption band.
A larger correction introduces a bump-like feature 
which is not coherent with the rest of the observations.
We explain the difficulty to correct better these 
spectra by the low level in the Chappuis bands, below the limit of sensitivity of the detector.
\section{Discussion} \label{dis}
\begin{figure*}[p]
\resizebox{!}{1.5\columnwidth}{\includegraphics{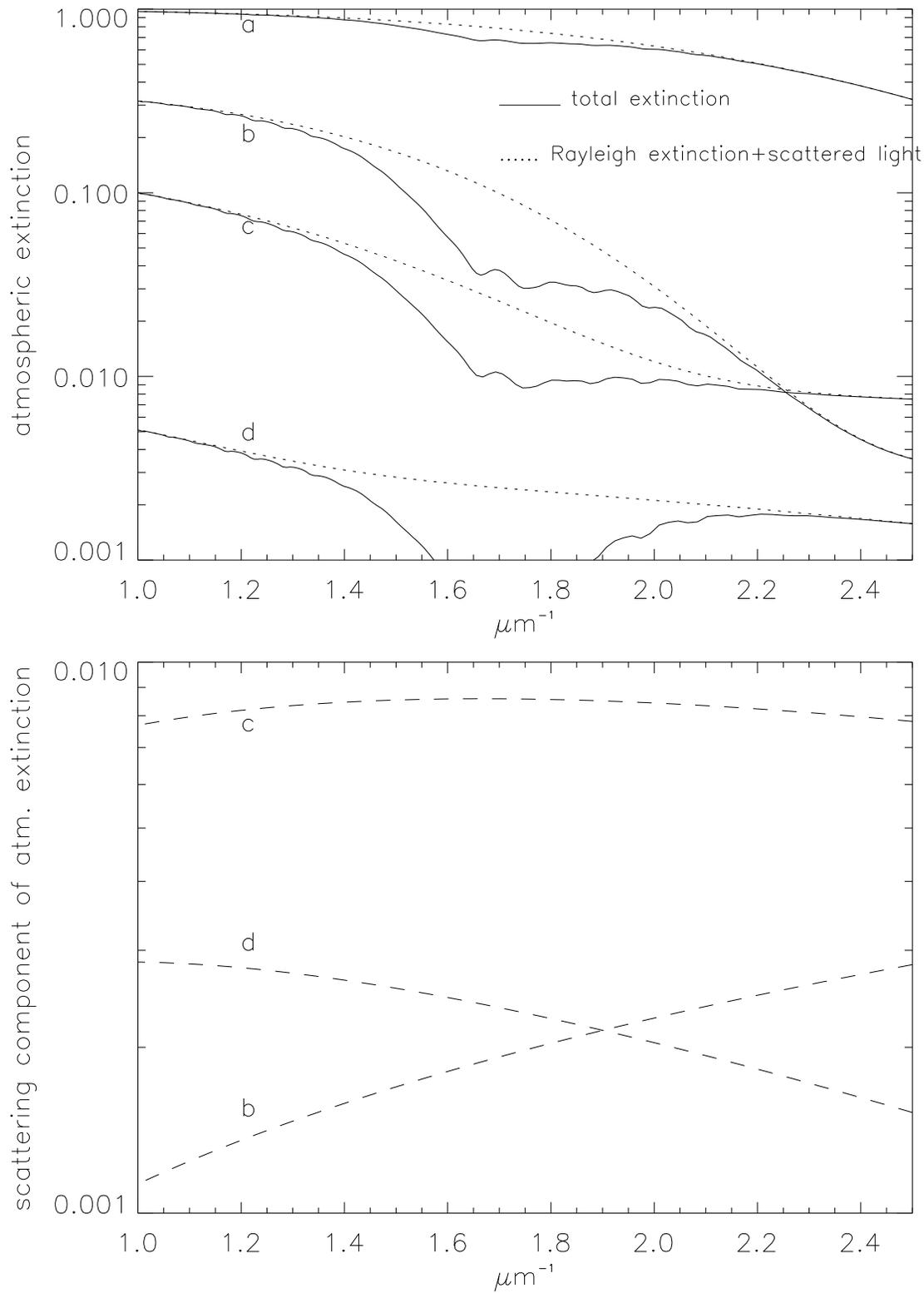}} 
\caption{
The four steps in atmospheric extinction (\emph{top})
and the forward scattering component alone (\emph{bottom}):
(a)=gas extinction alone, (b)=gas extinction+ rising
scattered light, (c)=gas extinction+ second order scattered light,
(d)=scattered light alone.
The analytical formulae are given in Table~\ref{tbl:recap}, and 
discussed in section~\ref{int}.
}
\label{fig:recap}
\end{figure*}
\subsection{Interpretation of the fits} \label{int}
\begin{table*}[]
\caption[]{Analytical expressions of the fits used in 
Figure~\ref{fig:recap}}		
       \[
    \begin{tabular}{|c|c|c|}
\hline
n$^{\circ}\,^{\,(1)}$ & spectrum$\,^{\,(2)}$ & Fit$\,^{\,(3)}$ \\   \hline
a &  3 &  $e^{-8\,10^{19}\sigma_{\lambda}}e^{-0.029/\lambda^4}$ \\
b &  5 & $0.4e^{-2\,10^{20}\sigma_{\lambda}}(e^{-0.16/\lambda^4}+3\,10^{-3}/\lambda)$\\
c &  7 & 
$0.2e^{-2\,10^{20}\sigma_{\lambda}}(e^{-0.17/\lambda^4}+7\,10^{-2}¥/\lambda)e^{-0.6/\lambda}$ \\
d &  10 & 
$0.01e^{-3\,10^{20}\sigma_{\lambda}}(e^{-0.5/\lambda^4}+0.8/\lambda)e^{-1.03/\lambda}$  \\
\hline
\end{tabular}    
    \]
\begin{list}{}{}
\item[$1$] Letters refers to the curves of Figure~\ref{fig:recap}. 
\item[$2$] Number of the spectrum in Table~\ref{tbl:spec}. 
\item[$3$] $\sigma_{\lambda}$ is the ozone absorption cross-section. 
\end{list}
\label{tbl:recap}
\end{table*}
The conclusions reached from the discussion of the spectra
(section~\ref{ana}) agree with the observational 
indications in our possession, the position of the sun which determines the optical 
path of sunrays in the atmosphere, the color index which confirms 
the presence of scattered light and measures its importance.

The intensity of the radiation field at any point of the atmosphere, 
before complete darkness, and in good atmospheric conditions (no heavy 
clouds), normalized by the spectrum of the sun, should 
be represented by one of the functions of Table~\ref{tbl:recap} 
(constants in these functions should be considered as orders of magnitudes), and illustrated 
by Figure~\ref{fig:recap}.

The spectra choosen for Figure~\ref{fig:recap} are the fit of 
spectra (3),  (5), (7) and (10).
Curve (a) is extinction by the gas alone.
The dotted line is the spectrum corrected for ozone absorption, and 
fits the $1/\lambda^{4}$ Rayleigh gas extinction.
Curve (b) is gas extinction plus the low optical depth rise of 
scattered light by aerosols.
In curve (c), the scattered component is displaced towards the longest 
wavelengths, and is perceived over all the visible spectrum.
The intensity of the scattered light does not vary much throughout the 
spectrum, because the maximum is reached at $1/\lambda\sim 
1.65\,\rm\mu m^{-1}$ ($\lambda\sim 6000\,\rm\AA$).
In spectrum (d), direct sunlight is a minor component of the radiation field;
the maximum of the scattered light has moved to the infrared.

In addition to the extinction of sunlight, there is a progressive diminution 
of the radiation field corresponding to the constant factor in front of each 
fit of Table~\ref{tbl:recap}.
This is because of the sunset, only the upper (and decreasing 
with time) part of the sun
participates to the radiation field at the cloud location.  
This also explains why the normalized intensity of the maximum of the 
scattered light of spectrum (d) 
is diminished compared to spectra (c) (bottom plot of 
Figure~\ref{fig:recap}).

The general form of the normalised radiation field measured at the 
balloon location is: 
$\propto (e^{-c_{2}¥/\lambda^{4}}+c_{4}¥/\lambda )e^{-c_{3}¥/\lambda}$.
It is not: $\propto (1+c_{4}¥/\lambda )e^{-c_{3}¥/\lambda}e^{-c_{2}¥/\lambda^{4}}$.
Which means that, scattered sunlight by aerosols behaves as if it was not 
extinguished by the gas.
No satisfying reason can yet be advanced to understand this particularity.
\subsection{The radiation field in the earth's atmosphere} \label{rad}
The four steps described in section~\ref{ana} correspond to 
a sun hidden behind successive, and thicker layers of the atmosphere.
Although the detector receives the light from all directions, the 
radiation field is (within the sensitivity of the instrument), all 
direct sunlight, and forward scattered light by aerosols:
the radiation field at the receptor location is dominated by light 
coming from the direction of the sun.
Due to the efficiency of forward scattering by aerosols, the 
radiation field in the atmosphere is still, at the time the sun 
disappears below the horizon, oriented by the direction of the sun.

In the stratosphere, extinction is of Rayleigh type (regardless of the absorption bands), 
most of which must be due to nitrogen ($80\%$ of the gas).
The light scattered by the gas in the stratosphere and in the 
troposphere, and received by the detector, should vary as 
$\propto \lambda^{-4}\,e^{-a/\lambda^{4}¥}$ \citep{rayleigh}.
It is weak compared to direct sunlight since it is not detected;
isotropic scattering, although it comes from a much larger angle, is 
much less efficient than forward scattering. 

The time scale for noticeable 
variations in the spectrum of the radiation field is of order of a 
minute (Table~\ref{tbl:spec}).
\subsection{The atmosphere viewed from earth} \label{terre}
On earth, an observer looking  in a different direction from the 
sun receives the light scattered by particles in the 
atmosphere.
If the direction of observation is far enough from 
the sun, forward scattering is  not efficient, because the angle of 
scattering is large.
The light received by the observer must be light isotropically scattered by molecules in the atmosphere.
If the radiation field in the atmosphere is homogenous enough along 
the direction of observation, the spectrum of the light received on 
earth, normalised by the spectrum of the sun, must be proportional to $1/\lambda^{4}$ times 
one of the spectra of Figure~\ref{fig:recap}.
\subsection{Extinction theory} \label{ext}
The analysis given so far was based on the simple 
principles of the extinction theory.
Ozone absorption aside, the size of the particles
is the determining parameter to understand the spectrum of a sunset, 
regardless of the particular composition of the particles present in the atmosphere.

From the analytical fits of the spectra 
(Figure~\ref{fig:recap}) direct sunlight, 
and forward scattered sunlight can be separated.
The intensity of the scattered light, integrated over all directions, 
is (bottom plot of Figure~\ref{fig:recap}) between $1 \%$ and $5\%$ 
(if we scale the constant factor of fit $(c)$ to 1)  of the direct sunlight 
corrected for extinction.
This value is of the same order as the one estimated in \citet{uv4} for 
interstellar grains.
\subsection{Application to observations in astronomy} 
\label{astro}
An object observed through the atmosphere should be extinguished in 
the same way as sunlight is.
Since the light received from a far away object crosses both the 
stratosphere and the troposphere, the applied correction 
should have one of the general expression (b), (c), or (d), of 
Table~\ref{tbl:recap}.

The proportion of molecular and aerosol
extinctions along the line of sight is a crucial point to determine 
how far towards the infrared atmospheric extinction will be felt.
In thin atmospheres, the effect of aerosols under $1/\lambda 
=2\,\rm \mu m^{-1}¥$ ($\lambda\ge 5000\,\rm \AA$) should be negligible.
If the ozone column density is larger than $10^{19}\,\rm cm^{-2}$,
the shape of the spectrum can be modified substantially in 
the visible portion of the spectra, from 
$1/\lambda =1.5\,\mu \rm m^{-1}¥$ to  $1/\lambda =2.1\,\mu \rm m^{-1}¥$
($5000\,\rm \AA\le \lambda \le 7000\,\rm \AA$).

The beam of ground-based telescopes is much smaller  (e.g. less than a 
few arcsecond) than the field of view of SAOZ's conical miror.
Therefore, depending on the 
phase function of aerosols, the importance of scattered light (the 
$1/\lambda$ term) may be reduced.
If the stratospheric layer is of low optical depth, or can be 
considered constant between the direction of observation and the 
direction of the reference object used to calibrate the spectra, gas 
extinction (the $e^{-c_{2}/\lambda^{4}¥}$ term) tends to vanish; 
the $e^{-c_{3}/\lambda}$ aerosol extinction term is the only one to remain.
In this case, only the slope of the spectrum is modified.

The time-scale deduced from Table~\ref{tbl:spec} for noticeable variations of the atmospheric 
extinction (due to the rotation of earth) is of order one minute.
If the duration of the observation is of a few minutes or more,
the correction for atmospheric extinction may have a 
more complicated analytical form than the simple functions of Table~\ref{tbl:recap}.
In this case, it is not evident that atmospheric extinction can easily 
be removed.
Spectral observations of nebulae, which generally require a few 
minutes observing time, are first concerned.

Contamination by direct starlight scattered in the 
atmosphere is another source of uncertainty for spectral observations 
of nebulae.
Since forward scattering is governed by the angle of 
scattering and the number of scatterers, rather than by the distance (source of illumination)-(scattering 
medium) \citep{o1opt}, there may be a competition between the light scattered by the 
nebula and the light scattered in the atmosphere.
This effect can be estimated in several ways:
through the comparison of aerosols and nebular column densities, 
and by observing a non-reddened star similar 
to the star illuminating the nebula. 
The effect of atmospheric extinction on astronomical 
spectral observations, can also be investigated by observing the 
same objects at different air-masses, and on different days. 
\section{Conclusion} \label{conc}
The observation of a sunset by the SAOZ balloon experience has produced a 
succession of spectra of the radiation field at the balloon location.
We have isolated four characteristic kinds of spectra from this sequence, 
which we have fitted and analyzed.
The physical interpretation of each fit is in good agreement with the 
independent information in our possession on the path followed by sunrays through the 
atmosphere, between the sun and the balloon. 

Our first task in this analysis was to correct the spectra for the 
ozone absorption.
The corrected spectra are then analyzed as follows.

When the sun is observed through the stratospheric layer only, sunlight is 
attenuated by a factor $e^{-c_{2}/\lambda^{4}}$:
atmospheric extinction is mainly Rayleigh scattering by the gas (nitrogen essentially).

Coinciding with the entrance of sunrays into the troposphere, gas extinction is no 
longer enough to fit the spectra towards the UV.
A $c_{4}/\lambda$ term must be added.
This term corresponds to forward scattering of sunlight by large 
particles (aerosols) in the troposphere. 

On increasing the optical depth, both gas and aerosols extinctions increase.
Extinction by aerosols now introduces a $e^{-c_{3}/\lambda¥}$ term 
which affects scattered light and -but to a lesser extent- direct 
sunlight. 
Scattered light increases when the sun reaches the 
horizon; and is progressively perceived towards the larger wavelengths. 

When the sun is below the horizon, just before complete darkness, forward 
scattererd light dominates the whole visible spectrum.

We deduce that in daylight, and during sunset,  most of the light received at  the balloon 
location, and similarly at any point of the 
atmosphere, comes from the direction of the sun.
This radiation field is then the sum of 
two effects on sunlight.
One effect is extinction of the direct sunlight, mainly due to the gas.
This includes the broadband ozone absorption, and Rayleigh extinction 
by the gas 
with an extinction optical depth $\propto \lambda ^{-4}$.
This direct and extinguished sunlight component of the radiation 
field is the only one to consider if sunrays cross the stratosphere 
alone.
The second component of the radiation field is forward scattered sunlight by 
aerosols in the troposphere.
The spectrum of the forward scattered light by aerosols
is well fitted by a function $\propto \lambda^{-1}e^{-c_{3}¥/\lambda}$.
Sunlight scattered by aerosols progressively replaces direct 
extinguished sunlight at sunset, with an important diminution of the intensity 
of the radiation field. Scattered sunlight is at most a few percent of 
non-extinguished direct sunlight.

The spectrum of the light received at the earth's surface from a direction far 
enough from the sun should be proportional to $1/\lambda^{4}$ 
times the spectrum of the atmosphere on the line of sight (section~\ref{terre}).

Our analysis made use of the general principles of 
scattering theory, and did not require the use of any model.
We show that, without calling to deterministic models 
of the atmosphere, it is 
possible to understand, separate, and quantify the components of the 
radiation field, direct and scattered sunlights.

Since starlight should be extinguished by the atmosphere as sunlight is, 
we have (section~\ref{astro}) 
discussed applications of this work to 
ground-based observations in astronomy.
{}
\end{document}